# Kvantový Zenonův jev aneb Co nesejde z očí, nezestárne

Mikhail Lemeshko a Břetislav Friedrich

*Fritz-Haber-Institut der Max-Planck-Gesellschaft*
Faradayweg 6, D-14195 Berlín, Německo

## 1. Úvod

Podle jedné z proslulých Zenonových aporií se letící šíp ve skutečnosti nepohybuje. V Aristotelově drkotavé formulaci[1]: "jestliže vše to, co zaujímá stejný prostor, je v klidu, a jestliže to, co se pohybuje, zujímá takový prostor v každém okamžiku, je letící šíp v klidu." Ač se o vyvrácení této a ostatních Zenonových aporií postaral již Aristoteles sám a po něm diferenciální počet, brouk, kterého Zenon nasadil lidstvu do hlavy, způsobuje jistou míru trápení dodnes. Kupodivu nejen mezi filosofy.

Na Zenona z Eleje si v roce 1977 vzpomněla dvojice matematických fyziků, Misra a Sudarshan, kteří tehdy analyzovali problém časového vývoje nestabilních kvantových systémů, speciálně radioaktivního rozpadu atomového jádra.[2] Tato analýza ukázala, že spontánní přechod kvantového systému z nestabilního do stabilního stavu lze zastavit nepřetržitým pozorováním! Aniž by podali bližší vysvětlení toho, jak to vlastně myslí, Misra a Sudarshan začali tuto podivuhodnost nazývat kvantovým Zenónovým jevem. Název se ujal, ač se většina těch, kteří se tímto jevem zabývají, shoduje na tom, že jeho podstatu lépe vystihuje méně erudovaně znějící rčení "a watched pot never boils."[3]

Kvantový Zenonův jev souvisí s jedním z nejexotičtějších pojmů kvantové mechaniky, tzv. *kolapsem vlnové funkce*.[4] Pojem kolapsu vlnové funkce zachycuje skutečnost, že opakované měření provedené na kvantovém systému vede ke stejnému výsledku: měření představuje událost, která přeruší spojitý časový vývoj kvantové soustavy, daný časově závislou Schrödingerovou rovnicí, a "vybere" ze souboru možných konečných stavů stav jediný. Kvantový Zenonův jev pak spočívá v tom, že opakované měření kolaps vlnové funkce *zpomaluje* a tedy oddaluje "výběr" onoho konečného stavu. V případě spojitého měření je kolaps vlnové funkce zpomalen natolik, že k němu (za idealizovaných podmínek) nedojde vůbec.

Kvantový Zenonův jev byl poprvé experimentálně prokázán ve Winelandově laboratoři (NIST, Boulder) v roce 1990.[5] Následovala řada dalších experimentů v různých laboratořích a na různých kontinentech.[6-9] Zatím poslední z nich byl proveden v Ketterleho laboratoři (MIT, Cambridge, MA) v roce 2006 s použitím Bose-Einsteinova kondenzátu.[10] V období od roku 1990 vznikla také řada teoretických studií, stimulovaných stávajícími experimenty a stimulující experimenty nové.[11]

Pro úplnost dodejme, že aspekty procesu rozpadu kvantového systému relevantní pro kvantový Zenonův jev byly poprvé analyzovány Chalfinem v roce 1957.[12] Jeho práce tehdy prošla téměř bez povšimnutí, možná také kvůli tomu, že její autor nezvolil k popisu svých výsledků imaginativní terminologii.

Pokračující zájem o kvantový Zenonův jev vychází jak z fundamentální fyziky (zejména teorie měření) tak i z aplikací. Ty zahrnují potlačení dekoherence v kvantových počítačích,[13] neutrínovou fyziku,[14] kontrolu rychlosti fotodisociace molekul,[15] a dokonce redukci dávek záření v roentgenové neutronové tomografii.[16] V tomto článku bychom chtěli poskytnout přehled o tom, co v posledním třicetiletí studium kvantového Zenónonova jevu přineslo.

## 2. Kde a jak kvantový Zenonův jev nastává?

Kvantový Zenonův jev nastává při *časovém vývoji* kvantové soustavy. Tento vývoj je dán časově závislou Schrödingerovou rovnicí. Popišme nyní touto rovnicí ten nejjednodušší možný systém schopný časového vývoje – tzv. dvouhladinový systém. Idealizovaným příkladem takového systému je atom, disponující pouhými dvěma energetickými hladinami, viz obr. 1. Systém má energii $E_1$, jestliže je ve stavu $|1\rangle$, a energii $E_2$, je-li ve stavu $|2\rangle$. Obecně je systém v čase $t$ ve stavu, jenž je koherentní superpozicí stavů $|1\rangle$ a $|2\rangle$,

$$|\text{obecný stav}\rangle = c_1(t)|1\rangle + c_2(t)|1\rangle$$

kde $c_1(t)$ a $c_2(t)$ jsou na čase závislé koeficienty udávající poměrné zastoupení stavů $|1\rangle$ a $|2\rangle$. Koeficienty $c_1(t)$ a $c_2(t)$ splňují v každém časovém okamžiku normovací podmínku $|c_1(t)|^2 + |c_2(t)|^2 = 1$, která zaručuje, že pravděpodobnost toho, že systém nalezneme v čase $t$ ve stavu $|1\rangle$ je $p_1(t) = |c_1(t)|^2$ a pravděpodobnost toho, že systém nalezneme v čase $t$ ve stavu $|2\rangle$ je $p_2(t) = |c_2(t)|^2 = 1 - p_1(t) = 1 - |c_1(t)|^2$. "Nalezením systému ve stavu $|1\rangle$ nebo $|2\rangle$" přitom myslíme to, že vlnová funkce, příslušející obecnému stavu systému, zkolabuje v *jediný* z možných konečných stavů, tj. $|\text{obecný stav}\rangle \to |1\rangle$ nebo $|\text{obecný stav}\rangle \to |2\rangle$. Jelikož jsou měření navzájem nezávislá (výsledek následného měření nezavisí na výsledku měření předešlého), je pravděpodobnost $p_1^{(n)}(t)$ nalezení systému ve stavu $|1\rangle$ po $n$ opakovaných měřeních dána vztahem $p_1^{(n)}(t) = [p_1(t)]^n$, kde $t$ značí časový interval mezi jednotlivými měřeními (provedení $n$ měření tedy trvá čas $nt \equiv T$). Podobně pravděpodobnost $p_2^{(n)}(t)$ nalezení systému ve stavu $|2\rangle$ po $n$ opakovaných měřeních dána vztahem $p_2^{(n)}(t) = [p_2(t)]^n$. Dosud uvedené bychom mohli nazvat "*kvantovou aritmetikou.*"

Předpokládejme, že v čase $t = 0$ je systém ve vzbuzeném stavu $|2\rangle$, tj. $p_2(t=0) = |c_2(t=0)|^2 = 1$ a tedy $p_1(t=0) = |c_1(t=0)|^2 = 0$. Použijme nyní výsledek *kvantové dynamiky*, obsažený v časově závislé Schrödingerově rovnici: z té vyplývá, že pro velmi krátký časový interval, kdy $t$ je téměř ale ne docela rovno nule, je koeficient $c_1(t)$ úměrný uplynulému času $t$, tj. $c_1(t \to 0) = at$, kde $a$ je konstanta úměrnosti. Závislost koeficientu $c_1(t)$ na čase je schematicky znázorněna na obr. 2.

To ale znamená, že pravděpodobnost toho, že se systém v čase $t \to 0$ stále ještě nachází ve stavu $|2\rangle$, je dána vztahem $p_2(t \to 0) = 1 - |c_1(t \to 0)|^2 = 1 - (at)^2$. Odtud ihned plyne, že pravděpodobnost nalezení soustavy v nestabilním stavu $|2\rangle$ při měření provedeném po uplynutí času $T \equiv nt$ je $p_2(T \to 0) = 1 - (aT)^2$. Jak se má tato pravděpodobnost k pravděpodobnosti $p_2^{(n)}(t)$ nalezení systému ve stavu $|2\rangle$ po $n$ měřeních pravidelně opakovaných vždy po uplynutí časovém intervalu $t$? Pro $t \to 0$ je tato pravděpodobnost dána vztahem $p_2^{(n)}(t)$ $= \left[1 - |c_1(t)|^2\right]^n = \left[p_2(t)\right]^n = \left[1 - |c_1(t)|^2\right]^n = \left[1 - (at)^2\right]^n \approx 1 - n(at)^2$. Po čase $T$ tedy máme $p_2^{(n)}(T) = \left[1 - (at)^2\right]^n \approx 1 - n(at)^2 = 1 - (aT)^2/n$, což pro nepřetržitě (spojitě) prováděná měření, $n \to \infty$, dává $p_2^{(n \to \infty)}(T) \to 1$.

Získali jsme tak vskutku pozoruhodný výsledek: při nepřetržitém pozorování se nestabilní systém vůbec nerozpadne! Právě v tom spočívá to, čemu se říká kvantový Zenonův jev.

Poznamenejme, že zpomalení rozpadu nestabilní kvantové soustavy nastává již pro $n = 2$: pravděpodobnost toho, že soustavu nalezneme ve stavu $|2\rangle$ po měření v mezičase $t$, následovaném měřením v čase $2t$, je $p_2^{(2)}(2t \to 0) \approx 1 - 2(at)^2$, zatímco pravděpodobnost toho, že systém bude ve stavu $|2\rangle$ po uplynutí celého času $2t$ je $p_2(2t \to 0) = 1 - 4(at)^2$.

Obr. 2 rovněž ukazuje, že po uplynutí delšího časového intervalu je koeficient $c_1(t)$ úměrný odmocnině času $t$, tj. $c_1(t) = (bt)^{1/2}$, kde $b$ je konstanta úměrnosti. V tomto případě je $p_2(t) = 1 - |c_1(t)|^2 = 1 - bt$ a kvantový Zenonův jev nenastává. Např. pro $n = 2$ máme $p_2^{(2)}(2t) = (1 - bt)^2 \approx 1 - 2bt$ a zároveň $p_2(2t) = 1 - 2bt$, tj. stejný výsledek, jako kdyby měření v mezičase *t* nebylo vůbec provedeno. Je tedy kvantový Zenonův jev spojen s časovou nelinearitou pravděpodobnosti přechodu z nestabilního do stabilního stavu. Čas, během kterého je tato pravděpodobnost kvadratická, se nazývá *Zenonovým časem*. Zenonův čas je obvykle velmi krátký; např. pro elektrický dipolový přechod v atomu vodíku ze vzbuzeného stavu 2p do základního stavu 1s jsou to zhruba 4 femtosekundy.[17] Ketterleho experiment pracuje s přechodem v atomech sodíku, jehož Zenonův čas dosahuje řádově mikrosekund. To je výhodou, která dovoluje získat vskutku spektakulárně přesná data o časovém vývoji systému a tedy i o kvantovém Zenonově jevu.

## 3. Experimenty prokazující kvantový Zenonův jev

Kvantový Zenonův jev zatím nebyl pozorován v případě jaderného rozpadu (jemuž byl původně ušit na míru). Důvodem je, že jaderný fragment (řekněme alfa částici) by bylo nutno pozorovat ve vzdálenostech od jádra odpovídajících řádově jaderným rozměrům. Jen tehdy by totiž pozorování mohlo proběhnout v Zenonově čase. Prostorové rozlišení stávajících detektorů na něco takového nestačí.

Kvantový Zenonův jev byl tedy pozorován při zářivých přechodech v atomových či iontových systémech. Experiment Winelandovy skupiny[5] byl proveden na souboru zhruba pěti set laserově ochlazených iontů berylia držených v Paulově pasti. Studován byl přechod mezi dvěma hyperjemnými hladinami, buzený radiofrekvenčním zářením. "Pozorování hrnce" se dělo prostřednictvím krátkých ultrafialových pulsů, excitujících Be$^+$ do třetího kvantového stavu, ze kterého ionty rychle relaxovaly do stavu základního. Tento přechod byl doprovázen vyzářením snadno detektovatelných fotonů, jejichž intensita vypovídala o obsazení základního stavu. V závislosti na tom, jak často ultrafialové pulsy dopadaly na soubor Be$^+$ v horním hyperjemném stavu, měnilo se jeho obsazení. Toto obsazení bylo možno jednoznačně dedukovat z měřeného obsazení základního stavu. Výsledky Winelandovy skupiny byly ve výtečném souhlase se shora popsanou teorií kvantového Zenonova jevu. Toschkova skupina[9] provedla podobná měření dokonce na jednotlivých iontech, rovněž zachycených v Paulově pasti. Souhlas byl opět zcela uspokojující. Obzvlášť důkladnou studii provedla nedávno Ketterleho skupina.[10]

Ketterle a spol. použili ve svém experimentu magneticky zachycený Bose-Einsteinův kondensát, sestávající zhruba z deseti tisíc atomů rubidia. Ten nechali pomalu oscilovat mezi základním stavem $|1\rangle$ a vzbuzeným stavem $|2\rangle$ s Rabiho frekvencí $\Omega$ (tyto oscilace byly indukovány kombinací mikrovlnného a radiofrekvenčního záření). *Pozorování* oscilujícího kondenzátu bylo uskutečněno tak, že byla měřena populace vzbuzeného stavu. K tomuto měření byla vyvinuta zvláštní varianta laserové absorpční spektroskopie: atomy ve stavu $|2\rangle$ byly vystaveny infračervenému laserovému paprsku, který je resonantně excitoval do vyššího vzbuzeného stavu. Přitom ty atomy, které absorbovaly infračervený resonantní foton, se absorpcí (s níž je spojen přenos hybnosti) translačně ohřály na teplotu 362 nK, tedy vysoko nad teplotu kondenzátu, která činila pouhých 15 nK. Tím přestaly být součástí kondenzátu. Po ukončení měření byly určeny populace stavů $|1\rangle$ a $|2\rangle$ v kondenzátu, na základě Sternovy-Gerlachovy separace obou stavů a následné balistické expanse kondenzátu po vypnutí magnetické pasti. Kvantový Zenonův jev byl studován jak v závislosti na frekvenci pulsů infračerveneho laseru, tak i na intensitě infračerveného laseru. Přitom byla použita též nulová frekvence, tj. spojitý parsek a tak tedy uskuteněno i spojité pozorování. Výsledky pulsního pozorování jsou ukázány na obr. 3, spolu s teoretickou křivkou. Shoda mezi experimentem a teorií je obdivuhodná. Povšimněme si, že role stavů $|1\rangle$ a $|2\rangle$ jsou zde zaměněny. To souvisí s tím, že studovaný indukovaný zářivý přechod se děje ze stavu $|1\rangle$ do stavu $|2\rangle$. Rekordní experimentálně určená pravděpodobnost toho, že systém přetrvá ve stavu $|1\rangle$ byla $p_1^{(n)}(T) = 0{,}984$, dosažená pro počet měření $n = 506$. Poločas zářivého přechodu byl přitom prodloužen téměř na dvousetnásobek převrácené Rabiho frekvence, $198 \times 1/\Omega$. Podobně přesvědčivá shoda mezi experimentem a teorií byla nalezena též v případě spojitého pozorování. Nezbývá tedy zřejmě nic jiného, než se s existencí kvantového Zenonova jevu vypořádat – a začít ho též využívat v aplikacích.

## 4. Aplikace

Kvantový Zenonův jev pomalu nachází použití jako spojenec v našem nerovném zápase s dekoherencí. Kvantová dekoherence nastává při interakci kvantového systému s klasickým okolím, a představuje ztrátu schopnosti kvantového systému vytvářet superpozici stavů a tedy se účastnit na interferenčních jevech. Právě těch je ale zapotřebí např. v kvantových počítačích, které pracují s kvantovou informací. Jednotkou kvantové informace je tzv. qubit, který je v kvantovém počítači realizován prostřednictvím koherentní superpozice dvojice stavů. Na rozdíl od bitu, jednotky klasické informace, která je buď jedničkou či nulou (podle toho zda elektrický proud teče nebo neteče) ale nikoli jejich lineární kombinací. Kvantový Zenonův jev dekoherenci zpomaluje.

Slibné je též např. použití v roentgenové a neutronové tomografii. Ukazuje se totiž, že kvantový Zenonův jev by mohl redukovat absorpci roentgenového či neutronového záření, a tak snížit zatížení vyšetřovaných tkání.[16]

*Tento článek věnujeme – s vřelým blahopřáním – Rudolfu Zahradníkovi k jeho velkému životnímu jubileu. Možná Rudolfova přetrvávající mladost a svěžest souvisejí i s tím, že byl sám během svého života hojně pozorován: do obratu v roce 1989 těmi, kteří se pozorováním snažili způsobit, aby Rudolf nenakazil mládež svou kultivovaností a vášní pro vědu a vůbec pro věci krásné a užitečné. To se ale nepodařilo. Mládež z vděčnosti Rudolfa infikovala svou mladostí. Chronicky.*

*Rudolf Zahradník je ovšem od obratu v roce 1989 také bedlivě sledován širší veřejností. Pro stejné vlastnosti jako dříve, ale s opačným cílem a ku prospěchu všech generací. Jaké potěšení se dívat!*

# Bibliografie

1. Aristoteles: Fyzika VI:9.

2. Misra, B., Sudarshan, E.C.G.: *18*, 756 (1977).

3. Parkinson, M.T.: Nucl. Phys. Ser. B *69*, 399 (1974).

4. Grifith, D.J.: Introduction to quantum mechanics (Prentice Hall, New Jersey, 1995), str. 381.

5. Itano, W.M., Heinzen, D.J., Bollinger, J.J., Wineland, D.J.: Phys. Rev. A *41*, 2295 (1990); viz též R. Pool, Science *246*, 888 (1989).

6. Fischer, M.C., Gutierrez-Medina, B., Raizen, M.G.: Phys. Rev. Lett. *87*, 0404021 (2001).

7. Toschek, P.E., Wunderlich, C.: Eur. Phys. J. D *14*, 387 (2001).

8. Balzer, C., Hannenmann, T., Reiss, D., Wunderlich, C., Neuhauser, W., Toschek, P.E.: Opt. Commun. *211*, 235 (2002).

9. Toschek, P.E.: Int. J. Mod. Phys. B *20*, 1513 (2006).

10. Streed, E.W., Mun, J., Boyd, M., Campbell, G.K., Medley, P., Ketterle, W., Pritchard, D.E.: Phys. Rev. Lett. *97*, 260402 (2006).

11. Koshino, K., Shimizu, A.: Phys. Rep. *412*, 191 (2005).

12. Chalfin, L.A.: Dokl. Akad. Nauk SSSR *115*, 227 (1957) (rusky); Khalfin, L.A.: Sov. Phys. JETP *6*, 1053 (1958).

13. Facchi, P., Tasaki, S., Pascazio, S., Nakazato, H., Tokuse, A., Lidar, D. A.: Phys. Rev. A *71*, 022302 (2005).

14. Boyanovsky, D., Ho, C.-.: J. High Energy Physics 7, 30 (2007); 10.1088/1126-6708/2007/07/030.

15. Prezhdo, O.V.: Phys. Rev. Lett. *85*, 4413 (2000).

16. Facchi, P., Hradil, Z., Krenn, G., Pascazio, S., Rehacek, J.: Phys. Rev. A *66*, 012110 (2002).
17. Facchi, P., Pascazio, S.: Phys. Lett. A *241*, 139 (1998).

**Popis k obrázkům**

Obrázek 1: Schéma dvouhladinového systému. Stavu $|1\rangle$ přísluší energie $E_1$, stavu $|2\rangle$ energie $E_2$. Obecně je systém ve stavu, jenž je koherentní superpozicí stavů $|1\rangle$ a $|2\rangle$. Při měření zkolabuje vlnová funkce obecného stavu systému buď ve stav $|1\rangle$ nebo ve stav $|2\rangle$.

Obrázek 2: Závislost rozvojového koeficientu $c_1(t)$ základního stavu $|1\rangle$ dvouhladinového systému na čase $t$. V případě klasického rozpadu je koeficient $c_1(t)$ úměrný veličině $[1 - \exp(-bt)]^{1/2}$ (tečkovaná křivka), která odpovídá exponenciálnímu rozdělení pravděpodobností rozpadu vzbuzeného stavu $|2\rangle$. V případě kvantového rozpadu je koeficient $c_1(t)$ úměrný veličině $[1 - \exp(-a^2 t^2)]^{1/2}$ (plná křivka), která odpovídá Gaussově rozdělení pravděpodobností rozpadu vzbuzeného stavu $|2\rangle$ pro velmi krátké časy. Zatímco pro $t \to 0$ je $[1 - \exp(-a^2 t^2)]^{1/2}$ přibližně rovno $at$ (čerchovaná křivka), pro delší čas $t$ přechází Gaussovo rozdělení v exponenciální. Kvantový Zenonův jev má svůj původ ve vztahu $c_1(t) = at$. Viz text.

Obrázek 3: Pravděpodobnost $p_1^{(n)}(T)$ přetrvání počátečního stavu v závislosti na počtu $n$ opakování měření během času $T$, kterého by bylo třeba k přenosu celé populace stavu $|1\rangle$ do stavu $|2\rangle$ v nepřítomnosti měření. Křivka odpovídá teoretické pravěpodobnosti toho, že systém přetrvá ve stavu $|1\rangle$ v případě $n$ ideálních měření, $p_1^{(n)}(T) = 1 - (aT)^2/n$, kde $aT = \pi/2$. Povšimněme si, že role stavů $|1\rangle$ a $|2\rangle$ jsou zde zaměněny. Viz text. Převzato z práce[10]: E.W. Streed, J. Mun, M. Boyd, G.K. Campbell, P. Medley, W. Ketterle, and D.E. Pritchard, "Continuous and Pulsed Quantum Zeno Effect," Phys. Rev. Lett. **97**, 260402 (2006).

**Quantum Zeno effect, or What's kept in sight, won't age**

We present the physics of the quantum Zeno effect, whose gist is often expressed by invoking the adage "a watched pot never boils." We review aspects of the theoretical and experimental work done on the effect since its inception in 1977, and mention some applications. We dedicate the article - with our very best wishes - to Rudolf Zahradnik at the occasion of his great jubilee. Perhaps Rudolf's lasting youthfulness and freshness are due to that he himself had been frequently observed throughout his life: until the turn-around in 1989 by those who wished, by their surveillance, to prevent Rudolf from spoiling the youth by his personal culture and his passion for science and things beautiful and useful in general. This attempt had failed. Out of gratitude, the youth has infected Rudolf with its youthfulness. Chronically. Since 1989, Rudolf has been closely watched by the public at large. For the same traits of his as before, but with the opposite goal and for the benefit of all generations. What a joy to keep him in sight!

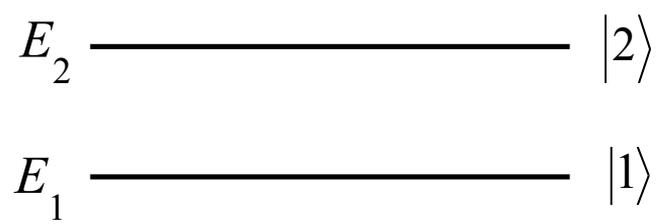

Fig. 1

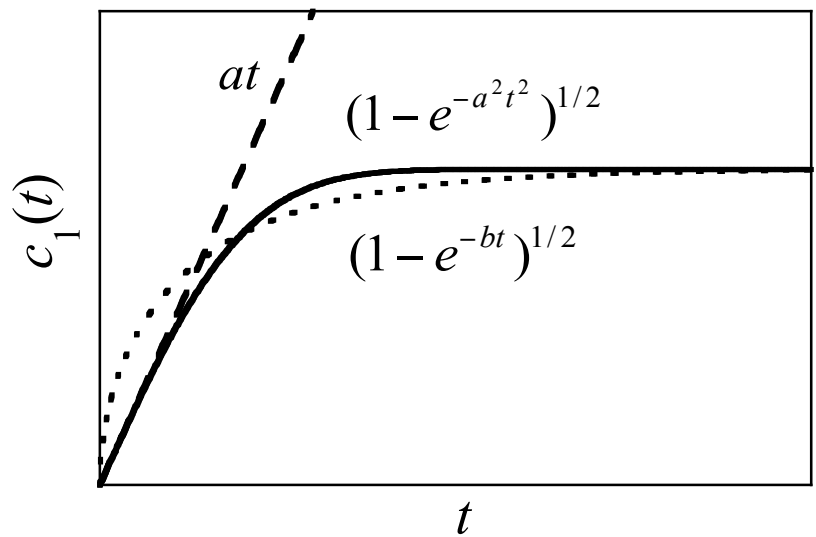

Fig. 2

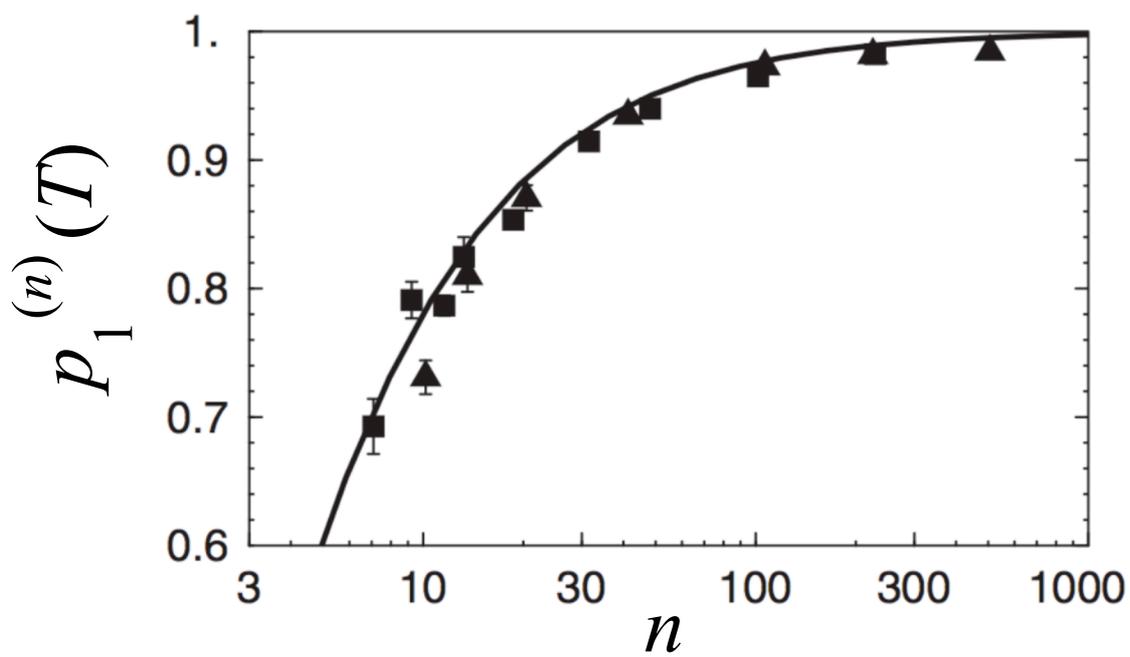

Fig. 3